\documentclass[pra,twocolumn,showpacs,preprintnumbers,amsmath,amssymb,nofootinbib]{revtex4} 
\usepackage{./figures}
\usepackage{./commands}

\usepackage{./biblio_aps}

\begin{document}

\title{Bose-Einstein Condensates  in Optical Quasicrystal Lattices}

%\author{------------- authors -------------}
\author{L.\ Sanchez-Palencia$^{1,2}$}
%  \email{}
%  \homepage{}
\author{L.\ Santos$^{3}$}
\affiliation{
$^1$Laboratoire Charles Fabry, Institut d'Optique,
Universit\'e Paris-Sud XI, F-91403 Orsay, France\\
$^2$Institut f\"ur Theoretische Physik, Universit\"at Hannover, D-30167 Hannover, Germany \\
$^3$Institut f\"ur Theoretische Physik III, Universit\"at Stuttgart, D-70569 Stuttgart, Germany}

\date{\today}

\begin{abstract}
We analyze the physics of Bose-Einstein condensates confined in 2D
quasi-periodic optical lattices, which offer an intermediate situation
between ordered and disordered systems. First, we analyze the
time-of-flight interference pattern that reveals quasi-periodic long-range
order. Second, we demonstrate localization effects associated with
quasi-disorder as well as quasiperiodic Bloch oscillations associated with
the extended nature of the wavefunction of a Bose-Einstein condensate in
an optical quasicrystal. In addition, we discuss in detail the crossover
between diffusive and localized regimes when the quasi-periodic potential
is switched on, as well as the effects of interactions.
\end{abstract}


\maketitle
%%%%%%%%%%%%%%%%%%%%%%%%%%%%%%%%%%%%%%%%%%%%%%%%%%

%%%%%%%%%%%%%%%%%%%%%%%%%%%%%%%%%%%%%%%%%%%%%%%%%%
%\section{Introduction}

% Optical lattices
The last few years have witnessed a fastly growing interest
on ultracold atomic gases in laser-generated periodic potentials 
(optical lattices, OLs). These
present neither defects nor phonons, offering
a powerful tool for investigating the quantum behavior of
periodic systems under unique control possibilities.
Thus, ultracold atoms trapped in OLs show 
fascinating resemblances with
solid-state physics,
which range from Bloch oscillations
\cite{dahan1997,morsch2001} and Wannier-Stark ladders
\cite{wannier96}, to Josephson arrays of
Bose-Einstein condensates (BECs) \cite{josephson}, or to
the superfluid to Mott-insulator transition \cite{mott1}.

% Non-standard optical lattices
These remarkable experiments have been performed in regular cubic OLs. 
However, these lattices do not exhaust the rich possibilities
offered by optical potentials. More sophisticated
lattice geometries have been proposed, as honey-comb
\cite{honeycomb} or Kagom\'e and triangular \cite{kagome} lattices.
Beyond, controlled defects may be introduced to generate random or
pseudorandom potentials \cite{horak1998}, allowing for the 
realization of Kondo-like physics \cite{Paredes},
Anderson localization and Bose-Glass phases \cite{disorder}. 
Exploiting this possibility, laser speckle fields have been employed 
very recently to produce BECs 
in random potentials \cite{inguscio,clement2005,schulte2005} 
opening very exciting new experimental possibilities. 

% Quasicrystals (general)
Bridging between ordered and disordered structures,
quasicrystals (QC) have attracted a wide interest since their discovery in
1984 \cite{shechtman1984}. QCs are a long-range ordered materials
but without 
translational invariance, and consequently they share properties
with both ordered crystals and amorphous solids \cite{quasicrystals}.
In particular, QCs show intriguing structure \cite{structure} as
well as electronic conduction 
properties \cite{conduction} at the border between ordered and disordered systems.

% Optical QCs
Surprisingly, up to now, only few works have been devoted to
optical analogues of QCs, despite of the fact that OLs 
offer dramatic possibilities for designing a
wide range of geometries \cite{grynberg2000}.
Optical QCs have been first studied in laser cooling experiments \cite{guidoni}, in which
the atomic gas, far from quantum degeneracy, 
was confined in a dissipative OL where quantum coherence was lost
due to spontaneous emission. 
In these systems, the temperature and spatial diffusion
were found to behave similarly as in periodic OLs.
The physics of 1D quasiperiodic OLs has also been
subject of recent research in the context of cold atomic gases,
including a proposal for the atom-optical
realization of Harper's model \cite{drese97}, and the analysis of
Fibonacci potentials \cite{eksioglu04}.

% Aim of this letter
In this Letter, we study the dynamics of a BEC 
in a 2D optical QC.
First, we show that the BEC wavefunction
displays quasi-periodic long-range order, a property that may be
easily probed via matter-wave interferometry.
Second, we show that macroscopic quantum coherence
dramatically modifies the transport on the lattice 
compared with the dissipative case.
On the one hand, due to quasi-disorder in optical QC, spatial
localization occurs, in contrast to ballistic expansion in periodic lattices.
The crossover between ballistic expansion and localization is analyzed when the quasi-periodicity
of the lattices is continuously increased.
On the other hand, we show that due to extended character of the BEC wavefunction,
Bloch oscillations take place. These oscillations are however quasi-periodic rather than periodic.
Additionally, we briefly discuss the effects of the interatomic interactions in the BEC diffusion.

%%%%%%%%%%%%%%%%%%%%%%%%%%%%%%%%%%%%%%%%%%%%%%%%%%
%\section{Our system}

%------------------------%
\begin{figure}[!ht]
\hspace{-.2cm}
\mbox{\epsfxsize 1.67in \epsfbox{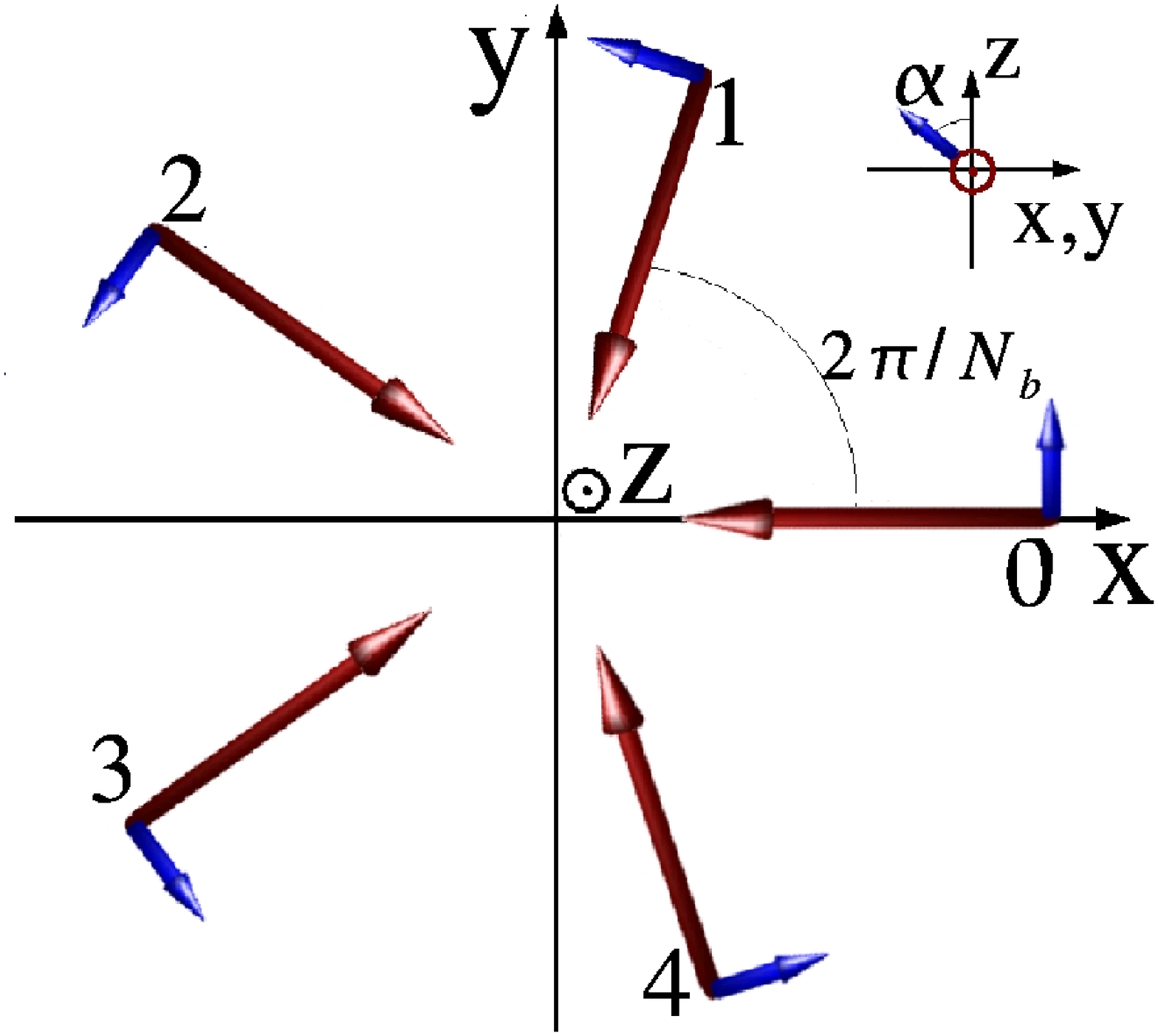}}
\hspace{.0cm}
\mbox{\epsfxsize 1.67in \epsfbox{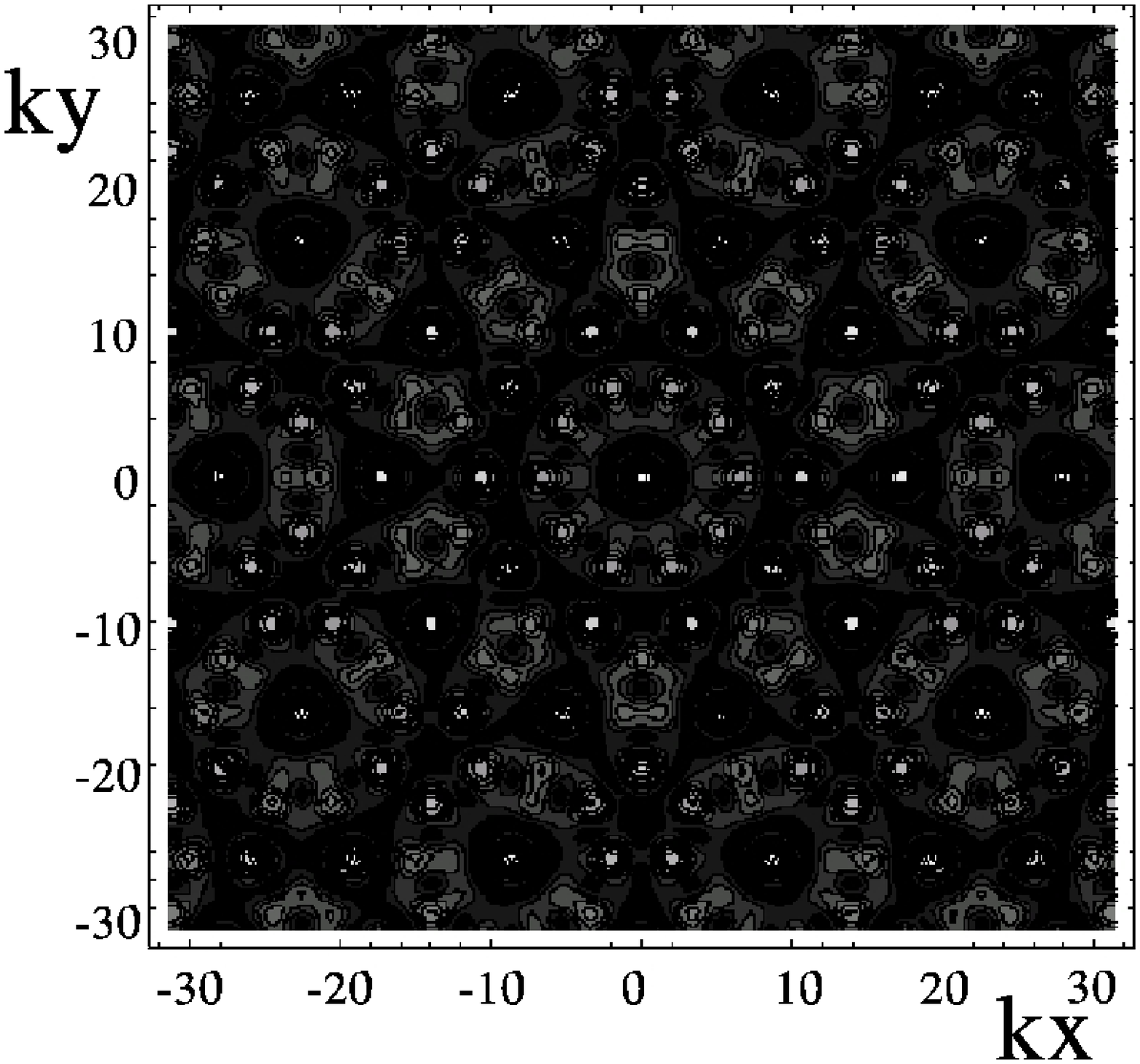}}
\caption{Left: laser arrangement (see text for details). 
Right: Quasi-periodic lattice potential for $N_\textrm{b}=5$, $V_0<0$ and
$\alpha_j=0$. White points correspond to potential minima.}
\label{fig:config}
\end{figure}
%------------------------%

In the following we consider a dilute Bose gas
trapped in the combination of a smooth harmonic potential
$V_\textrm{ho}(\vect{r})~=~\frac{M}{2}\left(\omega_\perp^2 \vect{r}_\perp^2+\omega_z^2 z^2\right)$
plus an OL $V_\textrm{latt}(\vect{r}_\perp)$. In the previous expression,
$M$ is the atomic mass, $\omega_j$ are the harmonic
trap frequencies, and $\vect{r}_\perp = (x,y)$ is the
position vector on the lattice plane. We assume $\omega_z$ to be large enough 
to keep a 2D physics on the $xy$ plane.
We consider a
laser configuration~\cite{guidoni} consisting on $N_\textrm{b}$
laser beams arranged on the  $xy$ plane with
$N_\textrm{b}$-fold
symmetry rotation (Fig.~\ref{fig:config}).
The polarization $\vect{\epsilon}_j$ of laser $j$ with
wavevector $\vect{k}_j$ is linear
and makes an angle $\alpha_j$ with the $xy$ plane.
The optical potential is thus \cite{grimm2000}:
%++++++++++++++++++++++++%
\be
V_\textrm{latt}(\vect{r}_\perp) =
\frac{V_0}{|\sum_j \mathcal{E}_j|^2} \left| \sum_{j=0}^{N_\textrm{b}}
\mathcal{E}_j \vect{\epsilon}_j ~
e^{-i (\vect{k}_j \cdot \vect{r}_\perp+\varphi_j)} \right|^2 ~,
\label{eq:lattice}
\ee
%++++++++++++++++++++++++%
where $0 \leq \mathcal{E}_j \leq 1$ stand for eventually different
laser intensities and $\varphi_j$ are the corresponding phases. In the following,
we are mostly interested in the fivefold symmetric
configuration ($N_\textrm{b}=5$, $\mathcal{E}_j=1$), similar to the
Penrose tiling \cite{penrose1974}, which supports no translational
invariance (see Fig.~\ref{fig:config}).
The lattice displays potential
wells which are clearly not periodically arranged.
We also consider the configuration obtained by switching off lasers
$1$ and $4$, which results in an anisotropic periodic lattice.

\begin{figure}[!ht]
\mbox{\epsfxsize 3.2in \epsfbox{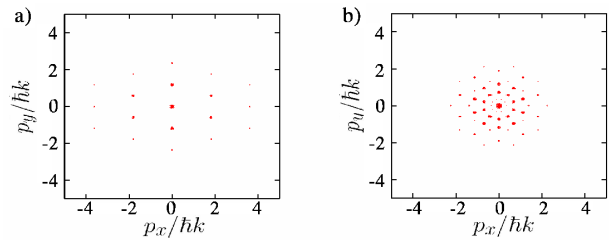}}
\caption{Matterwave interference pattern of a BEC released from
a combined OL and harmonic trap:
a) periodic case;
b) quasi-periodic case. Both correspond to $^{87}$Rb and $V_0=-10 E_\textrm{R}$,
where $E_\textrm{R}=\hbar^2 k^2/2M$ is the recoil energy.}
\label{fig:interference}
\end{figure}

%%%%%%%%%%%%%%%%%%%%%%%%%%%%%%%%%%%%%%%%%%%%%%%%%%
%\section{Matter wave interferometry}
\paragraph*{1) Equilibrium properties - }
The stationary BEC wavefunction $\psi_0$ is
obtained from evolution in imaginary time of the 2D Gross-Pitaevskii equation (GPE):
%++++++++++++++++++++++++%
\be
i \hbar \partial_t \psi = \left[ -\hbar^2 \vect{\nabla}^2/2M +
V_\textrm{ho} + V_\textrm{latt}
+ g_{2\textrm{D}} |\psi|^2 \right]
\psi ~,
\label{eq:GPE}
\ee
%++++++++++++++++++++++++%
where $g_{2\textrm{D}} = \sqrt{8\pi\hbar^3\omega_z/M}a_\textrm{sc}$, with 
$a_\textrm{sc}$ the $s$-wave scattering length.
In order to elucidate the long-range order properties of the BEC,
we compute the momentum distribution of $\psi_0$.
In the periodic case (Fig.~\ref{fig:interference}a), as expected, the momentum distribution
displays discrete peaks corresponding to combinations of
elementary basis vectors of the reciprocal lattice,
$n_1 \overrightarrow{\kappa}_1+n_2 \overrightarrow{\kappa}_2$
with integer coefficients $n_1$ and $n_2$. As obtained
in previous experimental works \cite{pedri2001,greiner2001}, this confirms the
periodic long-range order of $\psi_0$.
The quasi-periodic case (Fig.~\ref{fig:interference}b) is 
more intriguing, resulting in a more complex structure.
The momentum distribution also displays
sharp peaks, being the signature of 
a long-range order which is quasiperiodic rather than periodic \cite{note2}.
As in the periodic case,
the positions of the peaks are linear combinations of
integer numbers of $N_\textrm{b}=5$ wavevectors:
$\sum_{j=0}^{N_\textrm{b}-1}{n_j \overrightarrow{\kappa}_j}$
where $\overrightarrow{\kappa}_0=\overrightarrow{k}_1-\overrightarrow{k}_0$
and $\overrightarrow{\kappa}_j = \mathcal{R}[\phi_j]\overrightarrow{\kappa}_0$ is
the wavevector obtained by a rotation of angle
$\phi_j=2\pi j/N_\textrm{b}$ of $\overrightarrow{\kappa}_0$.
The reciprocal lattice thus clearly shows a five-fold
rotation symmetry incompatible with any translation invariance \cite{aschcroft}.
This resembles the Penrose tiling \cite{penrose1974}
and the first solide state QCs observed 
via Bragg diffraction \cite{shechtman1984}.

The discussed momentum distribution can be {\it directly imaged} 
via matter-wave interferometry after a time-of-flight expansion \cite{basdevant2002}.
Indeed, although the interactions are crucial for determining local
populations of each potential well, they do not contribute
significantly to the free BEC expansion after release from the trap \cite{pedri2001}.
Such measurements, standard in periodic OLs
\cite{pedri2001,greiner2001}, can be easily extended to quasi-periodic ones.

%------------------------%
%------------------------%

%%%%%%%%%%%%%%%%%%%%%%%%%%%%%%%%%%%%%%%%%%%%%%%%%%
%\section{Coherent diffusion of a BEC in a quasi-periodic lattice}
\paragraph*{2) Quantum transport - }
Certainly, not all physical
properties of optical QCs can be directly interpolated from the
behavior of periodic lattices. Indeed, solid QCs show
intriguing dynamical properties that are not yet completely understood
\cite{quasicrystals}. In the following, we investigate dynamical properties
of quasi-periodic lattices. 

\begin{figure}[!ht]
\infig{27em}{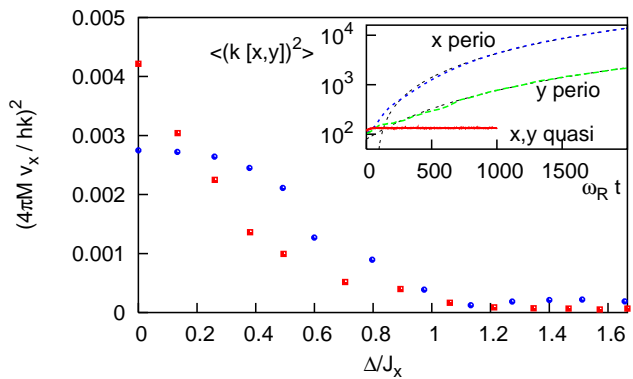}
\caption{
Crossover from ballistic to localization regimes with (squares) and without (circles) interactions, 
for $V_0=-7.5 E_\textrm{R}$. Inset: Coherent diffusion in periodic and quasi-periodic 
lattices for $V_0=-5 E_\textrm{R}$. 
Fits to $\langle r_j^2 (t) \rangle = r_{j 0}^2 +2D_{j}t+v_{j}^2t^2$ in the periodic case are also shown.
}
\label{fig:expan}
\end{figure}

\paragraph*{Coherent diffusion - }
Starting from the equilibrium wavefunction $\psi_0$, we consider the 
situation
in which the harmonic trap is switched-off at $t=0$, letting 
the BEC evolve in the OL. The BEC expansion is then computed using a Crank-Nicholson algorithm for the 
real time-dependent GPE~(\ref{eq:GPE}).
Figure~\ref{fig:expan} (inset) shows the time evolution of 
$\langle x^2 \rangle$ and $\langle y^2 \rangle$ of the interacting
BEC along $x$ and $y$
respectively.
In the periodic case, the condensate expands coherently as one expects from tunnel couplings
between adjacent lattice sites. 
In order to infer a convenient fitting functional for the expansion of the interacting 
BEC, let us recall that in free space for large times 
$\langle r_j^2 (t) \rangle \propto v_j^2 t^2$ with $v_j \propto 1/M$ \cite{basdevant2002}.
In periodic lattices, the inertia is enhanced, and the expansion is expected to be as 
in free space but substituting the atomic mass $M$ by an effective mass $M^* > M$. We thus expect
$v_j \propto 1/M^{*}$ and $M/M^* \propto J$ where $J$ is the site-to-site tunneling rate
\cite{kramer2002}.
Numerical computations for various depths 
of the lattice potential $V_0$ show that $v_j^2$ decreases exponentially with $V_0$ as expected 
from the well-known exponential decay of $J$.
Anisotropic ballistic expansion of the BEC reflects the anisotropy of our periodic lattice.

The behavior of the BEC in the quasiperiodic lattice is dramatically different, since 
after a short transient the BEC localizes \cite{footnote1}
(inset of Fig.~\ref{fig:expan}).
This behavior strongly contrasts with results obtained in the context of
laser cooling where similar classical normal expansion 
($\langle r_j^2(t) \rangle \sim 2 \tilde D_j t$) was found for both periodic and
quasi-periodic OLs \cite{guidoni}.
Here, spatial localization is a coherent effect induced by quasi-disorder due
to the lack of periodicity.
Indeed, the BEC populates localized (Wannier-like) states centered on each
lattice site. In the periodic case these states
have all the same energy and are strongly coupled through quantum tunneling. On the contrary, in the
quasi-periodic lattice, the sites have different energies. In particular, the typical 
difference of depths of adjacent sites (denoted $\Delta$ below) can be of the order of 
magnitude of (but smaller than) the potential depth. 
The tunneling is not resonant and the BEC localizes.

%phase transition
The remarkable flexibility of OLs \cite{grynberg2000} allows for the accurate
study of the competition between tunneling and quasi-disorder. 
By ramping up gradually the intensity of lasers $1$ and $4$ while keeping constant 
$0$, $2$ and $3$, one turns continuously from an anisotropic periodic lattice
to a five-fold symmetric quasiperiodic one, and hence from
ballistic expansion to spatial localization. 
For small intensity of the control lasers $1$ and $4$, the quasiperiodicity is
mainly {\it compositional} (the sites are still periodically displaced but the
on-site energies are different from site to site). We define the quasi-disorder $\Delta$
parameter as the variance of the differences of on-site energies in adjacent sites.
From the previous discussion and to compare to the results
of the non-degenerate case \cite{guidoni} we fit
%++++++++++++++++++++++++%
\be
\langle r_j^2 (t) \rangle = \langle r_{j0}^2 \rangle + 2D_{j}t + v_{j}^{2}t^2 ~.
\label{eq:fit}
\ee
%++++++++++++++++++++++++%
In the considered range of parameters, all calculations fit well with Eq.~(\ref{eq:fit})
with a negligible diffusive term $2D_{j}t$.
We characterize the expansion along the $x$-direction through the ballistic velocity $v_j$.
The behavior of $v_x$ versus the quasi-disorder parameter
$\Delta$ is shown in Fig.~\ref{fig:expan} for the interacting BEC and it is 
compared to the non-interacting case. For the latter, we simultaneously switched 
off the interactions at $t=0$ \cite{footnote2}. 
In both cases, as expected, coherent diffusion dramatically decreases when quasi-disorder 
increases. Spatial localization occurs for $\Delta \gtrsim J_x$, with  
$J_x$ the tunneling rate between adjacent sites along the $x$-direction \cite{footnote3}.
This supports the interpretation that competition between coherent 
tunneling and inhomogeneities turns into localization as soon as
tunneling becomes non-resonant. 

%------------------------%
%------------------------%

To understand the effect of interactions that can help ($\Delta \lesssim 0.2 J_x$ in Fig.\ref{fig:expan}) 
or hinder ($\Delta \gtrsim 0.2 J_x$ in Fig.\ref{fig:expan}) diffusion, note first that 
two phenomena contribute to localization: 
(i) initial inhomogeneities (due to disorder and harmonic confinement) that appear in the dynamics 
through the interaction term $g_{2D}|\psi_0|^2$ [see Eq.(\ref{eq:GPE})]
and (ii) inhomogeneities associated to quasi-disorder. 
Because of these inhomogeneities, quantum tunneling is not resonant and thus less efficient.
However, during diffusion, the interaction energy is converted into kinetic energy and
this tends to fasten the expansion. For small quasi-disorder, the second phenomenon dominates 
so that interactions contribute to expansion whereas for larger quasi-disorder, the inhomogenities
significantly hinder tunneling so that interactions contribute to localization.
The non-trivial interplay between disorder and interactions will be the subject of further research.

\begin{figure}[!ht]
\infig{27em}{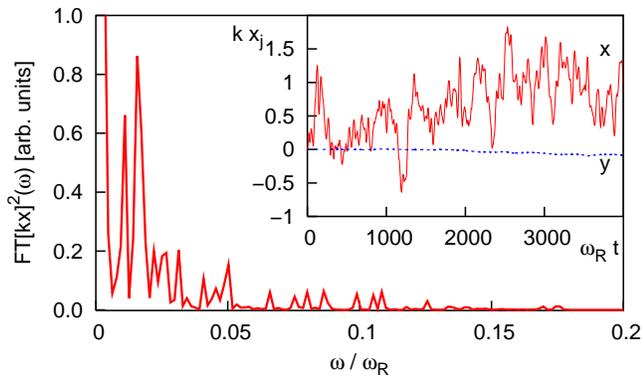}
\caption{Fourier transform of the mean position of a BEC in a periodic or quasiperiodic lattice.
Inset: Time evolution of a BEC in a tilted quasiperiodic lattice. All beams have the same intensity, 
$V_0=-2 E_\textrm{R}$ and 
$V_\textrm{tilt}=0.002 E_\textrm{R} \times kx$.}
\label{fig:bloch}
\end{figure}

\paragraph*{Quasiperiodic Bloch oscillations - }
One of the most appealing predictions of the quantum theory 
of solids \cite{aschcroft} is that homogeneous
static forces induce oscillatory rather than constant motion in 
periodic structures \cite{blochosc}.
The corresponding Bloch oscillations have already been 
observed in superlattice superconductors \cite{bloch1993}
and on cold atoms in OLs \cite{dahan1997,morsch2001}. 
It is a fundamental question whether such a phenomenon also exists in 
less ordered systems like QCs.
Arguments based on general spectral properties of QCs 
\cite{janot2000} and numerical simulations of 1D Fibonacci lattices 
\cite{diez1996} support the existence of Bloch oscillations in
quasiperiodic lattices. However, to the best of our knowledge, 
this effect has never been observed experimentally.
Using accelerated lattices \cite{dahan1997,morsch2001} or 
gravity \cite{roati2004} (we consider the latter), this question can
be addressed experimentally in the discussed arrangement 
(Fig.~\ref{fig:config}). 
Starting from $\psi_0$ we switched-off 
the harmonic trap and the interactions at time $t=0$ and tilt the
quasiperiodic lattice in the $x$ direction \cite{footnote4}. 
The latter evolution of the quantum gas is shown in the inset 
of Fig.~\ref{fig:bloch}. We find noisy-like oscillatory motion 
in the (tilted) $x$ direction and no motion 
in the (non-tilted) $y$ direction. The oscillations in the 
$x$ direction are clearly not periodic \cite{footnote5}. However, 
they definitely have an ordered structure, which is evidenced by the appearance 
of discrete sharp peaks in the time Fourier transform of the BEC mean position 
$\langle x(t) \rangle$ (Fig.~\ref{fig:bloch}), corresponding 
to a quasiperiodic motion \cite{note2}.

%------------------------%
%------------------------%

The Bloch-like quasiperiodic oscillations can be interpreted as follows.
Both in periodic and quasiperiodic lattices, the BEC wavefunction extends 
over many lattice wells, and can be decomposed into a sum of localized 
(Wannier-like) states. Due to the applied external force, these energy states 
are arranged in a Wannier-Stark ladder. In periodic lattices, 
the energy separation $\Delta E$ between the ladder states is fixed, 
leading to periodic Bloch oscillations of period $\propto \hbar/\Delta E$. However, 
for quasiperiodic lattices, a discrete set of different (non commensurate) 
differences of on-site energies in adjacent wells occurs (i.e. a non-equally 
spaced Wannier-Stark ladder) leading to quasiperiodic (instead of periodic) 
oscillations. Purely random potentials would result in a continuous 
set of differences of on-site energies leading, as expected, to the 
disapearance of any sort of Bloch-oscillations.

%%%%%%%%%%%%%%%%%%%%%%%%%%%%%%%%%%%%%%%%%%%%%%%%%%
%\section{Conclusion}
Summarizing, we have investigated the physics of BECs trapped in optical QCs.
We have shown that (i) the equilibrium BEC wavefunction displays long-range quasiperiodic
order and that (ii) quantum transport shares properties with both ordered and disordered systems.
On the one hand, because of quasi-disordered inhomogeneities, diffusion turns from ballistic to localization
when quasi-periodicity is switched on. On the other hand, because of coherence extending over several 
lattice sites, quasiperiodic Bloch oscillations occur in quasiperiodic BECs.

The discussed arrangement can be easily generated using standard techniques, offering 
an exciting tool for controlled studies of the transition between periodic, quasiperiodic and 
fully disordered systems in cold gases, a major topic of current experimental research \cite{inguscio,clement2005,schulte2005}.
In addition, the system can be used to address experimentally some
unsolved issues on QCs, as e.g. the application of the renormalization 
theory to the 2D case \cite{quasicrystals}.

%%%%%%%%%%%%%%%%%%%%%%%%%%%%%%%%%%%%%%%%%%%%%%%%%%
% \vspace{1.cm}
% \paragraph*{} \acknowledgements
We thank P. Verkerk, M. Inguscio, P. Pedri and H. Fehrmann for
discussions.
We acknowledge support from the Deutsche Forschungsgemeinschaft
(SFB 407 and SPP1116), RTN Cold Quantum Gases,
ESF PESC BEC2000+, Humboldt Foundation and CNRS.

%%%%%%%%%%%%%%%%%%%%%%%%%%%%%%%%%%%%%%%%%%%%%%%%%%

\end{document}